\documentclass[9pt,twocolumn,twoside]{opticajnl}
\journal{opticajournal} 

\setboolean{shortarticle}{false}

\usepackage{lineno}
\usepackage{bm}
\usepackage{subcaption}
\usepackage{nicefrac}
\usepackage{mathrsfs}
\usepackage{amsmath,amssymb,amsthm}
\usepackage{graphicx}
\usepackage{mathtools}

\author[1*]{Ehsan Faghihifar}
\affil[1]{Independent Researcher, Tehran, Iran}
\affil[*]{eh.faghihi@gmail.com}

\title{Bounds on Propagation Constants in Dielectric Gratings}

\begin{abstract}
We establish analytical bounds on the normalized longitudinal propagation constants $\beta=k_z/k_0$ of modes in lossless dielectric diffraction gratings. If $\epsilon_{\max}$ denotes the maximum relative permittivity, we show that regular modes ($\beta^2\in\mathbb R$) satisfy $\beta^2\le\epsilon_{\max}$ despite the lack of a scalar dispersion relation. We further show that the so-called \emph{ghost} modes ($\beta^2\notin\mathbb R$), admitted by the non-Hermitian nature of the associated eigenproblem, satisfy $|\operatorname{Re}(\beta)|\le\sqrt{\epsilon_{\max}}/2$. Both bounds are obtained directly from the Fourier-series formulation of the fully coupled vectorial eigenproblem.
\end{abstract}

\setboolean{displaycopyright}{false} 

\begin{document}

\maketitle

\section{Introduction}
Gratings considered in this work are transversely periodic, longitudinally invariant optical structures that are isotropic, non-magnetic, and lossless, with permittivity profile $\epsilon(x,y)$ defined over the unit cell $\Omega$. Longitudinal propagation in a grating leads to an eigenproblem whose eigenvalues are the squared longitudinal propagation constants and whose eigenfunctions are the modes, forming the basis of modal methods such as the analytical modal method (AMM)~\cite{Botten1981a, Botten1981b, Botten1981c, Botten1985, Sheng1982, Li1993b} and Fourier modal method (FMM)~\cite{Moharam1986, Moharam1982, Moharam1983, Moharam1995a, Li1997, Li1999, Li2001Chap} for solving far-field diffraction problems. The focus of this work is the eigenproblem itself, whose structural properties---selfadjointness, indefiniteness, and associated spectral features such as reality, completeness, distribution, and boundedness of the spectrum---govern both the spectral structure of grating propagation problems and the numerical methods used to solve them.

The spectral theory of Sturm--Liouville (SL) operators provides the classical framework underlying modal descriptions of 1D periodic (i.e., lamellar) gratings in TE/TM settings, where selfadjointness ensures real, discrete spectra and modal completeness~\cite{Zettl2005Book, Weidmann1987Book}. In  non-selfadjoint or indefinite settings, arising from electromagnetic losses or metallic constituents, these guarantees break down: eigenvalues may become complex, classical spectral bounds may fail, and completeness requires biorthogonal or generalized spectral systems~\cite{Mingarelli1982, Behrndt2013a, Curgus1989, Kocabas2009, Botten1981b, Hanson2013Book, Naimark1968Book}. While small losses act as a perturbation, a sign change in permittivity induces a qualitatively different spectral regime. These spectral anomalies, rooted in the indefinite nature of the problem, are well documented in TM metallic gratings~\cite{Li1993a, Foresti2006, Popov2004a}.

For 2D periodic (i.e., crossed) gratings, the governing equation is fully vectorial and the absence of a well-established differential framework and analytical solutions makes their spectral properties less tractable. Crucially, the resulting matrix eigenproblem is non-Hermitian even for lossless dielectric gratings, so non-real eigenvalues are not excluded a priori. In 1D such behavior arises only when the permittivity changes sign, which suggests that spectral phenomena documented in TM metallic gratings may have counterparts here without any metallic constituent. This study is concerned solely with spectral bounds for this general case. As the first result, we show that all regular modes of lossless dielectric gratings satisfy $\beta^2\le\epsilon_{\max}$, where $\epsilon_{\max}=\sup_{\Omega}\epsilon(x,y)$. The bound follows directly from the finite Fourier representation of the periodic propagation eigenproblem, without recourse to averaging or effective-medium arguments. In TM metallic lamellar gratings, propagation constants exceeding the bound set by the maximum permittivity have been reported~\cite{Tishchenko2005, Foresti2006}, showing that homogenization-based intuition is not a reliable guide in this context.

Our second result concerns \emph{ghost modes}, a spectral phenomenon arising from the non-Hermitian structure of the eigenproblem and so termed here by analogy with complex ghost states in indefinite Sturm--Liouville theory~\cite{Mingarelli2011}. Such solutions are characterized by non-real squared propagation constants even in lossless settings~\cite{Foresti2006, Sheng1982, Sturman2007a, Sturman2007b}. These modes are known to play a critical role in the convergence of modal expansions in TM metallic lamellar gratings~\cite{Foresti2006}. For ghost modes in lossless dielectric gratings, we establish the bound $|\operatorname{Re}(\beta)|\le\sqrt{\epsilon_{\max}}/2$; to the best of our knowledge, no comparable bound has previously been reported for dielectric gratings of general geometry. Such bounds are properties of the propagation spectrum itself, holding at every truncation order and hence for the modes they approximate; within this Fourier formulation, a computed eigenvalue violating them is erroneous.

The formulation employed here is a strong-form Fourier spectral discretization of Maxwell's equations in second-order form, as in FMM and related coupled-wave approaches. The bounds are shown to hold under Fourier factorization~\cite{Li1997} as well, and therefore across the formulations used in practice. The spectral relation of the resulting matrix eigenproblems to the underlying operator is a separate question, with rigorous convergence results available for broad classes of problems~\cite{Chatelin1981, Chatelin2011Book}, including discontinuous-coefficient differential eigenproblems in variational settings~\cite{Babuvska1978}. Beyond its numerical role, the Fourier formulation yields exact analytical bounds on the propagation spectrum of lossless dielectric gratings of arbitrary geometry through elementary linear algebra, for a fully coupled vectorial problem to which classical Sturm--Liouville theory does not directly apply.

\section{Formulation and Preliminaries}

Let the permittivity be piecewise smooth with $0\!<\!\epsilon_{\min}\!\le\!\epsilon(x,y)\!\le\!\epsilon_{\max}$ over $\Omega \!=\! [0, {\lambda}_0\ell_x] \!\times\! [0, {\lambda}_0\ell_y]$, with ${\lambda}_0$ the vacuum wavelength. With the time-harmonic convention $\exp(i\omega t)$ and $k_0 \!=\! 2\pi/{\lambda}_0\!=\!\omega/c$ the vacuum wavenumber, the wave equation for the electric field $\bm{E}(\bm{r}) \!=\! \left(\bm{E}_t(x,y) + E_z(x,y) \hat{\bm{z}}\right)\,\exp({-i k_z z})$ can be expressed as
\begin{equation}\label{eq:wave}
\left(\nabla_t - i k_z \hat{\bm{z}}\right)\times \left(\nabla_t - i k_z \hat{\bm{z}}\right) \times \bm{E} 
= k_0^2 \epsilon \bm{E},
\end{equation}
where $\nabla_t = \partial_x \hat{\bm{x}} + \partial_y \hat{\bm{y}}$ is the transverse gradient operator.   
By Floquet--Bloch theory, fixing the Floquet multiplier turns the continuum of solutions for $k_z$ into a discrete set of eigenvalues.
Using $\nabla\cdot(\epsilon\bm{E})\!=\!0$ to rewrite \eqref{eq:wave} in terms of the transverse component only, and proceeding analogously for the magnetic field from its dual wave equation, one obtains
\begin{align}
\nabla_t^\perp\!\left(\nabla_t^\perp\cdot\bm{E}_t\right)
+\nabla_t\!\left(\epsilon^{-1}\nabla_t\cdot(\epsilon\bm{E}_t)\right)
+k_0^2\epsilon\,\bm{E}_t &= k_0^2\beta^2\bm{E}_t, \label{eq:operator-et}\\
\nabla_t\!\left(\nabla_t\cdot\bm{H}_t\right)
+\epsilon\nabla_t^\perp\!\left(\epsilon^{-1}\nabla_t^\perp\cdot\bm{H}_t\right)
+k_0^2\epsilon\,\bm{H}_t &= k_0^2\beta^2\bm{H}_t, \label{eq:operator-ht}
\end{align}
where $\nabla_t^\perp := -\partial_y \hat{\bm{x}} + \partial_x \hat{\bm{y}}$, $\beta := k_z/k_0$, and $\bm{H}_t$ is the transverse magnetic field component.

We seek a strong-form Fourier discretization of a differential equation with $\Omega$-periodic coefficients. The solutions are consequently semi-periodic by Floquet--Bloch theory. In a general setting, let $0<f_{\min}\le f(x,y)\le f_{\max}$ be such a coefficient, and let $g(x,y)$ be a semi-periodic function on $\Omega$.
Define $\mathbb{N}_n\!:=\!\{1,2,...,n\}$ and $\mathbb{Z}_n\!:=\!\{-n,...,n\}$, for any positive integer $n$, and let $M$ and $N$ denote truncation orders with respect to $x$ and $y$, and $S\!:=\!(2M+1)(2N+1)$ the total Fourier order. 

The truncation applies to $g$, whose Fourier coefficients are sought over $\mathbb{Z}_M\times\mathbb{Z}_N$; the coefficients of $f$ are unrestricted, but only those over $\mathbb{Z}_{2M}\times\mathbb{Z}_{2N}$ ever contribute. Let $s$ denote a properly defined bijective 2D ordering of the lattice points, mapping $\mathbb{Z}_M\times\mathbb{Z}_N$ onto $\mathbb{N}_S$ and defined on $\mathbb{Z}_{2M}\times\mathbb{Z}_{2N}$ for $f$. A truncated Fourier series representation of $f$ and $g$ can then be expressed as
\begin{equation}\label{eq:fourier-forms}
f=\sum\nolimits_{\,s} \bar{f}_{s}\,e^{-i{k_0}\tilde{\bm{k}}_{s} \cdot {\bm{r}_t}},\qquad
g=\sum\nolimits_{\,s} \bar{g}_{s}\,e^{-i{k_0}(\tilde{\bm{k}}_{s}+\tilde{\bm{k}}_0) \cdot \bm{r}_t},
\end{equation}
with $\bar{f}_{s}, \bar{g}_{s}$ Fourier coefficients, $\bm{r}_t\!=\!(x,y)$, $\tilde{\bm{k}}_s\!:=\!(\tilde{k}_{xs},\tilde{k}_{ys})\!=\!(m/\ell_x,n/\ell_y)$ the normalized transverse wave vectors, and $\tilde{\bm{k}}_0\!:=\!(\tilde{k}_{x0},\tilde{k}_{y0})$ the normalized Floquet wavevector, held constant throughout. Define the generalized $S\times S$ Toeplitz matrix $\operatorname{T}(f)$ and the $S\times 1$ column vector $\operatorname{V}(g)$, associated with the functions $f$ and $g$ described above, as follows:
\begin{equation}\label{eq:TV-def}
[\operatorname{T}(f)]_{s,s'} := {\bar{f}_{s-s'}},\qquad
[\operatorname{V}(g)]_s := {\bar{g}_{s}},\qquad
s,s'\in\mathbb{N}_S,
\end{equation}
where $s-s'$ denotes the index of the lattice difference, i.e., $s(m,n)-s(m',n')\!:=\!s(m-m',\,n-n')$. Note that the ordering is not unique; however, the standard choice $s(m,n)\!=\!(2N+1)(m+M)+{n+N+1}$ turns $\operatorname{T}$ into a block Toeplitz matrix of Toeplitz blocks (BTTB).
Moreover, we construct the following $S\!\times\!S$ diagonal matrices:
\begin{equation}\label{eq:N-def}
\bm{N}_u := \operatorname{diag}\left\{\tilde{k}_{us}+\tilde{k}_{u0}\right\}_s,\qquad  u\in\{x,y\}.
\end{equation}
Differentiation is diagonal on the truncated basis, yielding
\begin{equation}\label{eq:V-derivative}
\operatorname{V}(\partial_u g)=-ik_0\bm{N}_u\operatorname{V}(g),\qquad u\in\{x,y\}.
\end{equation}

Unlike derivatives, products are not explicit on a truncated basis. A finite Fourier series approximation of $fg$ with minimum-$L^2$-error still requires all the Fourier coefficients of $f$ and $g$; hence, $\operatorname{V}(fg)$ is not determined by $\operatorname{V}(f)$ and $\operatorname{V}(g)$ under any finite truncation, and must be approximated, incurring truncated convolution errors (TCE)~\cite{Faghihifar2022}. The most direct scheme takes
\begin{equation}\label{eq:TV-properties}
\operatorname{V}(fg)\approx\operatorname{T}(f)\operatorname{V}(g),
\end{equation}
the truncated form of the Fourier convolution theorem. Such an approximation is not unique: writing $h=f^{-1}g$ gives $\operatorname{V}(h) \!\approx\! \operatorname{T}(f^{-1})\operatorname{V}(g)$, while writing it as $fh=g$ gives $\operatorname{T}(f)\operatorname{V}(h)\!\approx\!\operatorname{V}(g)$, and hence, $\operatorname{V}(h)\!\approx\!\operatorname{T}(f)^{-1}\operatorname{V}(g)$. While $\operatorname{T}(f^{-1})$ and $\operatorname{T}(f)^{-1}$ are only asymptotically equal, they can be used interchangeably in our argument.

Recall that for a Hermitian matrix $\bm{M}$ and a nonzero vector $\bm{v}$, the Rayleigh quotient $R(\bm{M},\bm{v})\!:=\!\bm{v}^*\bm{M}\bm{v}/\|\bm{v}\|^2$ is a weighted average over the eigenvalues of $\bm{M}$. If $\lambda_{\min}$ and $\lambda_{\max}$ denote the minimum and maximum eigenvalues of $\bm{M}$, then
\begin{equation}\label{eq:R-properties}
\lambda_{\min}\le R(\bm{M},\cdot)\le\lambda_{\max},\qquad \lambda_{\max}^{\,-1}\le R(\bm{M}^{-1},\cdot)\le\lambda_{\min}^{\,-1},
\end{equation}
and by the min-max theorem the minima and maxima are met.

Now, let $g$ be the Fourier polynomial associated with $\bm{v}=\operatorname{V}(g)$ with $\tilde{\bm{k}}_0=\bm{0}$, so that $\|\bm{v}\|\!=\!\|g\|_2$ by Parseval's identity, where $\|g\|_2\!:=\!(\tfrac{1}{|\Omega|}\int_\Omega|g|^2\,d^2 r_t)^{1/2}$ denotes the $L^2$-norm over the cell. Expanding $ f|g|^2= g^*fg$ in terms of their Fourier series and averaging over the cell annihilates all but the zeroth-order term, yielding
\begin{equation}\label{eq:T-quadratic}
\frac{1}{|\Omega|}\int_\Omega f\,|g|^2\,d^2 r_t =\sum\nolimits_{s,s'} \bar{g}_{s}^{*}\,\bar{f}_{s-s'}\,\bar{g}_{s'} = \bm{v}^{*}\operatorname{T}(f)\bm{v},
\end{equation}
so that $f_{\min} \le R(\operatorname{T}(f),\bm{v}) \le f_{\max}$. Since the bounds in \eqref{eq:R-properties} are attained,
\begin{equation}\label{eq:T-lambda-bounds}
0<f_{\min}\le\lambda_{\min}(\operatorname{T})\le\lambda_{\max}(\operatorname{T})\le f_{\max},
\end{equation}
and in particular $\operatorname{T}(f)$ is positive definite.

\section{Regular Modes}

Now, let us denote for $u\in\{x,y\}$
\begin{equation}\label{eq:F-eu-def}
\bm{F} := \operatorname{T}(\epsilon),\qquad
\bm{e}_u := \operatorname{V}(E_{u}).
\end{equation}
We further define $\bm{e}_t \!:=\! [\bm{e}_x\,;\, \bm{e}_y]$, $\bm{N} \!:=\! [\bm{N}_x\,;\, \bm{N}_y]$, $\bm{N}_\perp \!:=\! [-\bm{N}_y\,;\, \bm{N}_x]$, $\bm{N}_0^2\!:=\!\bm{N}_x^2+\bm{N}_y^2$, and $\bm{F}_2 \!:=\! \operatorname{diag}(\bm{F},\bm{F})$, where the semicolon denotes vertical concatenation. Note that $\bm{N}_0$ is uniquely specified with real entries. We also assume $\bm{I}$ and $\bm{I}_2$ denote the $S\times S$ and $2S\times 2S$ identity matrices, respectively. Now, using \eqref{eq:V-derivative} and \eqref{eq:TV-properties}, \eqref{eq:operator-et} can be discretized into the following matrix eigenvalue equation:
\begin{equation}\label{eq:eig-main}
(\bm{P}-\beta^2\bm{I}_2)\bm{e}_t = \bm{0},\qquad \bm{P} := \bm{F}_2 - \bm{N}_\perp\bm{N}_\perp^* - \bm{N}\bm{F}^{-1}\bm{N}^*\bm{F}_2,
\end{equation}
where we used the $\operatorname{T}(\epsilon)^{-1}$ scheme for multiplication by $\epsilon^{-1}$, although using $\operatorname{T}(\epsilon^{-1})$ leads to the same result. Alternatives to \eqref{eq:eig-main} are more commonly derived from the Fourier expansion of Maxwell's equations \cite{Moharam1995a, Li1997}. Note that the bounds derived below are unconditional statements about the matrix $\bm{P}$ at each truncation order, and rely on no exact correspondence between \eqref{eq:eig-main} and \eqref{eq:operator-et}. The first thing to note about the modal matrix $\bm{P}$ is that it can simply be decomposed into
\begin{equation}\label{eq:AB-def}
\bm{P}=\bm{A}\bm{B},\qquad \bm{A} := \bm{I}_2 - \bm{N}\bm{F}^{-1}\bm{N}^*,\qquad \bm{B} := \bm{F}_2 - \bm{N}_\perp \bm{N}_\perp^*,
\end{equation}
which follows from the fact that $\bm{N}^*\bm{N}_\perp\!=\!\bm{N}_\perp^*\bm{N}\!=\!\bm{0}$. While $\bm{A}$ and $\bm{B}$ are Hermitian, they do not commute in general, so $\bm{P}$ is non-Hermitian. Nevertheless, while the eigenvalues of $\bm{P}$ and $\bm{P}^*$ are related by complex conjugation, $\bm{A}\bm{B}\!=\!\bm{P}$ and $\bm{B}\bm{A}\!=\!\bm{P}^*$ have the same eigenvalues. Consequently, the set of eigenvalues is closed under complex conjugation, comprising only real values and complex-conjugate pairs.

The decomposition also determines how modes relate to one another. If $\bm{A}\bm{B}\bm{e}_{t1}\!=\!\beta_1^2\bm{e}_{t1}$, then $\bm{e}_{t1}^*\bm{B}\bm{A}\!=\!\beta_1^{2*}\bm{e}_{t1}^*$, and multiplying from the right by $\bm{B}\bm{e}_{t2}$ gives $\beta_2^2\,\bm{e}_{t1}^*\bm{B}\bm{e}_{t2}\!=\!\beta_1^{2*}\bm{e}_{t1}^*\bm{B}\bm{e}_{t2}$. Hence any two modes with $\beta_1^{2*}\!\neq\!\beta_2^2$ satisfy $\bm{e}_{t1}^*\bm{B}\bm{e}_{t2}\!=\!0$: the modes are orthogonal not directly, but under the Hermitian form induced by $\bm{B}$, equivalently biorthogonal with respect to the adjoint problem $\bm{B}\bm{A}\bm{e}_t'\!=\!\beta^2\bm{e}_t'$ with $\bm{e}_t'\!=\!\bm{B}\bm{e}_t$.

By the same discretization applied to \eqref{eq:operator-ht}, the modal matrix for $\bm{h}_t \!:=\! [\operatorname{V}(H_x)\,;\,\operatorname{V}(H_y)]$ takes the dual form $(\tilde{\bm{P}}-\beta^2\bm{I}_2)\bm{h}_t=\bm{0}$, where
\begin{equation}\label{eq:AB-def-h}
\tilde{\bm{P}}:=\tilde{\bm{B}}\tilde{\bm{A}},\qquad
\tilde{\bm{B}}:=\bm{F}_2-\bm{N}\bm{N}^*,\qquad
\tilde{\bm{A}}:=\bm{I}_2-\bm{N}_\perp\bm{F}^{-1}\bm{N}_\perp^*,
\end{equation}
sharing the spectrum of \eqref{eq:eig-main}, both deriving from the same Maxwell system.

Now, we proceed by bounding the eigenvalues of \eqref{eq:eig-main}. First, assume $\beta^2\!>\!0$ is a positive real eigenvalue. The bound trivially holds for $\beta^2\!\le\! 0$. Expanding the left-hand side of \eqref{eq:eig-main} and using the fact that $\bm{A}\bm{N}_\perp\!=\!\bm{N}_\perp$, one can write
\begin{equation}\label{eq:eig-ord-simp}
\bm{A}\bm{B}\bm{e}_t = 
\bm{A} \bm{F}_2 \bm{e}_t - \bm{N}_\perp \bm{N}_\perp^* \bm{e}_t = \beta^2 \bm{e}_t,
\end{equation}
which can be rearranged into the following:
\begin{equation}\label{eq:eig-Cbeta}
\bm{A} \bm{F}_2 \bm{e}_t =  \bm{C}_{\beta} \beta^2 \bm{e}_t,\qquad \bm{C}_{\beta}:=\bm{I}_2 + \beta^{-2} \bm{N}_\perp \bm{N}_\perp^*.
\end{equation}
Note that $\bm{C}_{\beta}$ is Hermitian, since $\beta^2$ is real here and $\bm{N}_\perp\bm{N}_\perp^*$ is Hermitian by construction, with $R(\bm{C}_{\beta},\bm{v})=1+\beta^{-2}\|\bm{N}_\perp^*\bm{v}\|^2/\|\bm{v}\|^2>0$, which implies $\bm{C}_{\beta}$ is positive definite and hence invertible. Accordingly, the eigenvalue equation can be rearranged as follows, where $\bm{C}_{\beta}^{-1}$ is elaborated using Woodbury formula \cite{Horn2012Book}:
\begin{equation}\label{eq:eig-Cbeta-inv}
\bm{C}_{\beta}^{-1} \bm{A} \bm{F}_2 \bm{e}_t = \beta^2 \bm{e}_t,\;\quad\; 
\bm{C}_{\beta}^{-1}=\bm{I}_2 - \bm{N}_\perp \!\left( \beta^2 \bm{I} + \bm{N}_0^2 \right) ^{-1} \!\bm{N}_\perp^*.
\end{equation}
Since $\bm{C}_{\beta}$ is invertible, $\bm{C}_{\beta}^{-1}$ is Hermitian as well, and $\bm{A}$ commutes with it by $\bm{N}^*\bm{N}_\perp\!=\!\bm{N}_\perp^*\bm{N}\!=\!\bm{0}$, necessitating that their product is also Hermitian, denoted and expanded as follows:
\begin{equation}\label{eq:eig-H-def}
\begin{aligned}
\bm{H} := \bm{C}_{\beta}^{-1}\bm{A} 
&= \bm{A} - \bm{N}_\perp \! \left( \beta^2 \bm{I} + \bm{N}_0^2 \right) ^{-1}\!\bm{N}_\perp^*\\
&= \bm{I}_2 - \bm{N} \bm{F}^{-1} \bm{N}^* - \bm{N}_\perp \! \left( \beta^2 \bm{I} + \bm{N}_0^2 \right) ^{-1}\!\bm{N}_\perp^*.
\end{aligned}
\end{equation}
Now, we multiply both sides of \eqref{eq:eig-Cbeta-inv} on the left by $\bm{e}_t^*\bm{F}_2$ and rearrange to obtain
\begin{equation}\label{eq:eig-beta2-as-ratio}
\beta^2 = \frac{\bm{e}_t^* \bm{F}_2 \bm{H} \bm{F}_2 \bm{e}_t}{\bm{e}_t^* \bm{F}_2 \bm{e}_t} 
= \frac{R(\bm{H},\bm{F}_2\bm{e}_t)}{R(\bm{F}_2^{-1},\bm{F}_2\bm{e}_t)}.
\end{equation}
Note that \eqref{eq:T-lambda-bounds} and \eqref{eq:R-properties} imply $R(\bm{F}_2^{-1},\cdot)\ge \epsilon_{\max}^{\,-1}$. Moreover, from \eqref{eq:eig-H-def}, $\bm{H}-\bm{I}_2$ is negative semidefinite (negative definite if $\tilde{\bm{k}}_0\neq\bm{0}$), since for every vector $\bm{v}\neq\bm{0}$
\begin{equation}\label{eq:eig-H-neg-def}
\bm{v}^*(\bm{H}-\bm{I}_2)\bm{v} = - c_1\|\bm{N}^*\bm{v}\|^2 - c_2 \|\bm{N}_\perp^*\bm{v}\|^2\le 0,
\end{equation}
where $c_1=R(\bm{F}^{-1},\bm{N}^*\bm{v})$ and $c_2=R\left((\beta^2 \bm{I} + \bm{N}_0^2)^{-1},\bm{N}_\perp^*\bm{v}\right)$ are positive wherever defined, the corresponding term vanishing otherwise. 
This yields $R(\bm{H}-\bm{I}_2,\cdot)\!\le\! 0$ and subsequently, $R(\bm{H},\cdot)\!\le\! 1$. Having bounded the numerator and denominator in \eqref{eq:eig-beta2-as-ratio}, one obtains $\beta^2\!\le\! \epsilon_{\max}$, which concludes the first result of this paper. This bound is consistent with the asymptotic eigenvalue patterns of~\cite{Faghihifar2020}, which begin at $\epsilon_{\max}$ and decrease, though obtained there from semi-analytical estimates rather than proved.

\section{Ghost Modes}

For the second category, i.e., ghost modes, let $\beta^2\!=\!\rho+i\nu$ with $\nu\neq 0$. If $\operatorname{Re}(\beta)=0$, the bound is immediate, hence we assume
$\operatorname{Re}(\beta)\neq0$. Since $\beta^{2*}\!\neq\!\beta^2$, such a mode is self-orthogonal under $\bm{B}$, giving $\bm{e}_t^*\bm{B}\bm{e}_t \!=\! 0$, or $\bm{e}_t^*\bm{F}_2\bm{e}_t \!=\! \|\bm{N}_\perp^*\bm{e}_t\|^2$. Note from Maxwell's equations that $H_z\!=\!i(\omega\mu_0)^{-1}\left({\partial_x E_y}-{\partial_y E_x}\right)$, which translates into $\bm{h}_z\!=\!y_0\,\bm{N}_\perp^*\bm{e}_t$ for $\bm{h}_z \!=\! \operatorname{V}(H_z)$ and $y_0\!=\!\sqrt{\epsilon_0/\mu_0}$. Consequently, $\mu_0\,\|\bm{h}_z\|^2=\epsilon_0 \,\bm{e}_t^*\bm{F}_2\bm{e}_t$, or equivalently, for the Fourier polynomials $\bm{E}_t$ and $H_z$ associated with $\bm{e}_t$ and $\bm{h}_z$,
\begin{equation}\label{eq:ghost-balance}
\int_\Omega \epsilon_0\epsilon\, |\bm{E}_t|^2 \, d^2 r_t = \int_\Omega \mu_0 |H_z|^2 \, d^2 r_t.
\end{equation}
In the dual formulation, ghost modes likewise satisfy $\bm{h}_t^*\tilde{\bm{A}}\bm{h}_t\!=\!0$.

Now, we derive a bound governing the non-real $\beta^2$. Note that $\bm{C}_{\beta}$ of \eqref{eq:eig-Cbeta} is no longer Hermitian; however, it remains invertible. We prove this by contradiction. Singularity of $\bm{C}_{\beta}$ implies there exists a non-trivial vector $\bm{v}$ satisfying $\bm{C}_{\beta}\bm{v}=0$, giving $\bm{v}^*\bm{C}_{\beta}\bm{v}=0$. However, one can expand
\begin{equation}\label{eq:R-Cbeta-complex}
\bm{v}^*\bm{C}_{\beta}\bm{v} = \left( \|\bm{v}\|^2 + \rho|\beta|^{-4} \|\bm{N}_\perp^*\bm{v}\|^2 \right) 
- i\left( \nu |\beta|^{-4} \|\bm{N}_\perp^*\bm{v}\|^2 \right).
\end{equation}
Since $\nu\neq 0$, the vanishing of the imaginary part in \eqref{eq:R-Cbeta-complex} implies $\|\bm{N}_\perp^*\bm{v}\|\!=\!0$, which forces $\operatorname{Re}(\bm{v}^*\bm{C}_{\beta}\bm{v})\!=\!\|\bm{v}\|^2>0$, contradicting our initial assumption and yielding invertibility of $\bm{C}_{\beta}$. Now, rearranging \eqref{eq:eig-beta2-as-ratio} and replacing $\bm{H}$ from \eqref{eq:eig-H-def} implies
\begin{equation}\label{eq:beta2-complex}
\beta^2 \bm{e}_t^* \bm{F}_2 \bm{e}_t =
\bm{e}_t^*\bm{F}_2\bm{A}\bm{F}_2\bm{e}_t - \bm{e}_t^* \bm{F}_2 \bm{N}_\perp (\beta^2 \bm{I} + \bm{N}_0^2)^{-1} \bm{N}_\perp^* \bm{F}_2 \bm{e}_t.
\end{equation}
If we denote $n_s^2\!:=\!(\tilde{k}_{xs}+\tilde{k}_{x0})^2\!+\!(\tilde{k}_{ys}+\tilde{k}_{y0})^2$ for every index $s\in\mathbb{N}_S$, such that $\bm{N}_0^2\!=\!\operatorname{diag}\{n_s^2\}_s$, then $(\beta^2 \bm{I} + \bm{N}_0^2)^{-1} = \operatorname{diag}\left\{(\rho+n_s^2-i\nu)/|\beta^2+n_s^2|^2\right\}_s$. Replacing this in \eqref{eq:beta2-complex}, equating imaginary parts of both sides, and simplifying gives
\begin{equation}\label{eq:Mbeta-def}
\bm{e}_t^* \bm{F}_2 \bm{e}_t = \bm{e}_t^* \bm{F}_2 \bm{M}_{\beta} \bm{F}_2 \bm{e}_t,\quad 
\bm{M}_{\beta}\!:=\!\bm{N}_\perp \operatorname{diag}\left\{|\beta^2+n_s^2|^{-2}\right\}_s \bm{N}_\perp^*,
\end{equation}
where $\bm{M}_{\beta}$ is a Hermitian matrix. The above equation can equivalently be expressed as
\begin{equation}\label{eq:beta2-complex-imag-R}
R\left(\bm{F}_2^{-1}, \bm{F}_2 \bm{e}_t\right) = R\left(\bm{M}_{\beta}, \bm{F}_2 \bm{e}_t\right),
\end{equation}
where $\|\bm{F}_2 \bm{e}_t\|^2$ has been simplified from both sides, since $\bm{F}_2$ is positive definite and hence $\|\bm{F}_2 \bm{e}_t\| \neq 0$. To estimate the right-hand side of the equation above, we first note that $\bm{M}_\beta$ is at most half-rank, and shares all of its nontrivial eigenvalues with its cyclic permutation, given below,
\begin{equation}\label{eq:Mbeta-cycl}
\begin{aligned}
\bm{M}_{\beta}^{\,c}
&:=\operatorname{diag}\!\left\{|\beta^2+n_s^2|^{-2}\right\}_s \bm{N}_\perp^*\bm{N}_\perp \\
&= \operatorname{diag}\!\left\{|\beta^2+n_s^2|^{-2}\right\}_s \bm{N}_0^2 = \operatorname{diag}\!\left\{n_s^2|\beta^2+n_s^2|^{-2}\right\}_s,
\end{aligned}
\end{equation}
whose eigenvalues satisfy
\begin{equation}\label{eq:Mbeta-cycl-eig}
\begin{aligned}
\lambda_s(\bm{M}_{\beta}^{\,c})
&= \frac{n_s^2}{|\beta^2+n_s^2|^{2}} = \frac{n_s^2}{|\beta|^4+n_s^4+2\rho n_s^2}\\
&\le \frac{n_s^2}{2|\beta|^2 n_s^2+2\rho n_s^2} = \frac{1}{4\operatorname{Re}(\beta)^2}.
\end{aligned}
\end{equation}
Here, we used the fact that $|\beta|^4+n_s^4\ge 2|\beta|^2 n_s^2$ and $|\beta|^2+\rho = 2\operatorname{Re}(\beta)^2$. Since $\bm{M}_\beta$ is Hermitian positive semidefinite,
\begin{equation}\label{eq:R-Mbeta}
R\left(\bm{M}_{\beta},\bm{F}_2\bm{e}_t\right)
\le \lambda_{\max}(\bm{M}_{\beta})
= \lambda_{\max}(\bm{M}_{\beta}^{\,c})
\le \frac{1}{4\operatorname{Re}(\beta)^2}.
\end{equation}
Now, recall that for the left-hand side of \eqref{eq:beta2-complex-imag-R}, we showed earlier that $R(\bm{F}_2^{-1},\cdot)\ge \epsilon_{\max}^{\,-1}$. Putting this together with the result of \eqref{eq:R-Mbeta}, one may write
\begin{equation}\label{eq:beta2-complex-bound}
\epsilon_{\max}^{\,-1} \le R\left(\bm{F}_2^{-1}, \bm{F}_2 \bm{e}_t\right) = R\left(\bm{M}_{\beta}, \bm{F}_2 \bm{e}_t\right) \le \frac{1}{4\operatorname{Re}(\beta)^2},
\end{equation}
simplifying to $|\operatorname{Re}(\beta)|\!\le\!\sqrt{\epsilon_{\max}}/2$ and giving the intended bound for ghost modes.

\section{Scope of the Bounds}

We close with a few remarks. The derivations use the Fourier representation of $\epsilon$ only through two properties: that the matrix representing $\epsilon$ is Hermitian positive definite with $\lambda_{\max}\!\le\!\epsilon_{\max}$, and that the matrix representing $\epsilon^{-1}$ is Hermitian positive definite. By \eqref{eq:T-lambda-bounds}, both $\operatorname{T}(\epsilon)$ and $\operatorname{T}(\epsilon^{-1})^{-1}$ satisfy the required upper spectral bound, and both $\operatorname{T}(\epsilon)^{-1}$ and $\operatorname{T}(\epsilon^{-1})$ are Hermitian positive definite. The bounds therefore hold at every truncation order for any scheme whose $\epsilon$ and $\epsilon^{-1}$ matrices are of these types.

The same holds under Fourier factorization~\cite{Li1997}, as we now show. For $u\in\{x,y\}$, let $\operatorname{T}_u$ and $\operatorname{V}_u$ denote the one-dimensional counterparts of \eqref{eq:TV-def}, taken along $u$ alone with the trivial ordering, and acting entrywise on matrix- or vector-valued arguments, so that $\operatorname{T}_u(\bm{A})\!=\![\operatorname{T}_u(A_{ij})]$. With the BTTB ordering noted after \eqref{eq:TV-def} one has $\operatorname{T}\!=\!\operatorname{T}_x\!\circ\!\operatorname{T}_y$ and $\operatorname{V}\!=\!\operatorname{V}_x\!\circ\!\operatorname{V}_y$; the reversed composition corresponds to the reversed ordering, the two being related by a unitary permutation of the rows and columns.

Let $\bm{A}(u)$ be Hermitian, periodic, and bounded as $a_{\min}\!\le\!\lambda(\bm{A}(u))\!\le\! a_{\max}$ for every $u$, and let $\bm{G}(u)$ be the vector-valued Fourier polynomial associated with $\bm{v}\!=\!\operatorname{V}_u(\bm{G})$. Expanding $\bm{G}^*\bm{A}\bm{G}$ and averaging over the period annihilates all but the zeroth-order term, giving
\begin{equation}\label{eq:T-quadratic-matrix}
\frac{1}{\lambda_0\ell_u}\int_0^{\lambda_0\ell_u} \bm{G}^*\bm{A}\,\bm{G}\,du =\sum\nolimits_{n,n'} \bar{\bm{G}}_{n}^{*}\,\bar{\bm{A}}_{n-n'}\,\bar{\bm{G}}_{n'} = \bm{v}^{*}\operatorname{T}_u(\bm{A})\bm{v},
\end{equation}
so that $a_{\min}\!\le\!R(\operatorname{T}_u(\bm{A}),\bm{v})\!\le\! a_{\max}$ and, by the min-max theorem, the eigenvalues of $\operatorname{T}_u(\bm{A})$ lie in $[a_{\min},a_{\max}]$. The argument is that of \eqref{eq:T-quadratic}, one dimension at a time.

Li's formulation for the rectangular coordinate system leads to \eqref{eq:eig-main} with $\bm{F}_2$ alone replaced by $\operatorname{diag}(\bm{F}_{yx},\bm{F}_{xy})$, every other matrix being unchanged. Here $\bm{F}_{yx}\!=\!\operatorname{T}_y(\operatorname{T}_x(\epsilon^{-1})^{-1})$, and $\bm{F}_{xy}$ equals $\operatorname{T}_x(\operatorname{T}_y(\epsilon^{-1})^{-1})$ up to the unitary reordering above. For each fixed $y$, \eqref{eq:T-quadratic-matrix} with scalar $\bm{A}$ places the spectrum of $\operatorname{T}_x(\epsilon^{-1})$ in $[\epsilon_{\max}^{-1},\epsilon_{\min}^{-1}]$, so that of its inverse lies in $[\epsilon_{\min},\epsilon_{\max}]$; applying \eqref{eq:T-quadratic-matrix} again with $\bm{A}\!=\!\operatorname{T}_x(\epsilon^{-1})^{-1}$ preserves these bounds under $\operatorname{T}_y$, and likewise for $\bm{F}_{xy}$. Hence $\bm{F}_2$ remains Hermitian positive definite with $\lambda_{\max}\!\le\!\epsilon_{\max}$, and both bounds carry over at every truncation order.

The bounds also apply to 1D gratings, which in the dielectric case support no ghost modes. For the TM problem, for instance, the equation in terms of the magnetic field reads $\bm{F}\bm{A}_x\bm{h}_y\!=\!\beta^2\bm{h}_y$, with the Hermitian $\bm{A}_x\!:=\!\bm{I}-\bm{N}_x\bm{F}^{-1}\bm{N}_x$, whence $\bm{h}_y^*\bm{A}_x\bm{h}_y\!=\!\beta^2\bm{h}_y^*\bm{F}^{-1}\bm{h}_y$. The left-hand side is real, and $\bm{h}_y^*\bm{F}^{-1}\bm{h}_y>0$ since $\bm{F}$, and hence $\bm{F}^{-1}$, is positive definite for a dielectric grating, so $\beta^2$ must be real.

In TM metallic lamellar gratings, however, ghost modes are known to emerge~\cite{Foresti2006}, and must then satisfy $\bm{h}_y^*\bm{F}^{-1}\bm{h}_y\!=\!0$. This parallels the condition $\bm{h}_t^*\tilde{\bm{A}}\bm{h}_t\!=\!0$ found earlier, which arises without any sign change in permittivity. Propagation constants exceeding $\epsilon_{\max}$ are likewise reported in that setting~\cite{Tishchenko2005, Foresti2006}. In that case, the Fourier framework yields neither a bound nor boundedness itself. A boundedness result has nonetheless been established for a class of Sturm--Liouville problems that includes the TM metallic lamellar grating~\cite{Faghihifar2026}.

\section{Conclusion}

We have established two bounds on the propagation spectrum of lossless dielectric gratings of arbitrary geometry: $\beta^2\le\epsilon_{\max}$ for regular modes, and $|\operatorname{Re}(\beta)|\le\sqrt{\epsilon_{\max}}/2$ for ghost modes. Both follow from the algebraic structure of the Fourier-series formulation alone, hold at every truncation order, and are unaffected by the choice of Fourier factorization. The results rest on the dielectric assumption. Once the permittivity changes sign, the argument no longer applies, and the question of boundedness itself remains open for crossed gratings.


\begin{backmatter}
\bmsection{Disclosures} The author declares no conflicts of interest.
\bmsection{Data availability} No data were generated or analyzed in the presented research.
\end{backmatter}


\end{document}